\documentclass{article}

\usepackage{graphicx}
\usepackage{subfig}
\usepackage{enumitem}
\usepackage[hyphens]{url}
\usepackage{float}
\usepackage{spconf}
\usepackage{amsmath}

\title{Remap, warp and attend: Non-parallel many-to-many accent conversion with Normalizing Flows}

\name{\parbox{\linewidth}{\centering Abdelhamid Ezzerg$^{1}$ Thomas Merritt$^{1}$ Kayoko Yanagisawa$^{1}$ Piotr Bilinski$^{1}$ Magdalena Proszewska$^{2}$\sthanks{Work performed during an internship at Amazon.}  Kamil Pokora$^{1}$  Renard Korzeniowski$^{1}$  Roberto Barra-Chicote$^{1}$ Daniel Korzekwa$^{1}$}}
\address{\begin{tabular}{c}$^1$ Amazon Alexa $^2$ Jagiellonian University, Poland\end{tabular} \\
\textit{\{ezzerg, thommer, yakayoko, bilipiot, kamipoko, korenard, rchicote\}@amazon.com}}

\begin{document}

\ninept
\maketitle

\begin{abstract}
Regional accents of the same language affect not only how words are pronounced 
(i.e.,  phonetic content), but also impact prosodic aspects of speech such as speaking rate and 
intonation. This paper investigates a novel flow-based approach to accent conversion using normalizing flows. The proposed approach revolves around three steps: remapping the phonetic conditioning, to better match the target accent, warping the duration of the converted speech, to better suit the target phonemes, and an attention mechanism that implicitly aligns source and target speech sequences. The proposed remap-warp-attend system enables adaptation of both phonetic and prosodic aspects of speech while allowing for
source and converted speech signals to be of different lengths. Objective and subjective evaluations show that the proposed approach significantly outperforms a competitive  CopyCat baseline model in terms of similarity to the
 target accent, naturalness and intelligibility. 

\end{abstract}


\section{Introduction}
Speech attributes' conversion aims to generate synthetic speech from a source one by changing only attributes of interest. Speech attributes' conversion includes two notable tasks: Voice conversion (VC) ~\cite{Mohammadi2017An,Karlapati_2020,qian2019autovc,parrotron}, where the goal is to change the perceived speaker's identity while maintaining the linguistic and prosodical content of the source speech, and Accent Conversion (AC), whose goal is to generate new speech that sounds as if the source speaker was natively speaking the target accent. Accent conversion has a wide range of applications which include personalized voices for AI assistants and pronunciation feedback for language learners~\cite{motivation_1, motivation_2}.

There are two main families of models that convert speech attributes: parallel models 
and non-parallel models~\cite{lorenzotrueba2018voice}. 
Parallel methods for AC require a dataset of pairs of the  same utterance spoken in different accents. Such data is expensive to collect and difficult to obtain. Therefore, we focus in this paper on non-parallel approaches to accent conversion which only require transcribed speech.

Accent conversion approaches are varied in terms of the used tools and paradigms. Earlier AC methods relied on manipulating various features such as duration, intonation, pitch contours etc. using Hidden Markov Models (HMMs) ~\cite{accent_conversion_1,accent_conversion_6,accent_conversion_2, accent_conversion_5}. It is worth noting that these approaches were trained on parallel data and require access to a reference speech in the target accent at inference time. With the advent of more modern machine learning architectures, recent AC approaches started leveraging neural networks. Regardless of the used tools (DNNs or HMMs/GMMs), it is possible to distinguish multiple AC approaches in the literature. W. Li et al.~\cite{accent_conversion_4} leverages a Text-to-Speech (TTS) system for accent conversion in a two-stage training pipeline in order to create synthetic parallel data which is later used to train a conversion model. S. Aryal et al.~\cite{accent_conversion_3, accent_conversion_7} 
uses articulation data to improve accent conversion results; however, articulation data is difficult to collect and therefore 
limits the applications of this approach. Moreover, the approach is trained on a corpus of parallel data which further limits its applications.
Other approaches have tackled the AC problem using phonetic 
posteriograms~\cite{accent_conversion_2, accent_conversion_5}. While these approaches are non-parallel at inference time, they need a parallel training corpus.
A common aspect of existing approaches is that they assume the availability of parallel data~\cite{AC2021} or the availability of the source speech and a reference 
speech of the target accent at the time of inference~\cite{accent_conversion_4}. 
A notable exception to the previous observation is the work conducted by Z. Wang ~\cite{many-to-many-accent-conversion} where they tackle both VC and AC simultaneously. In their work, the author use an ASR system, in order to extract content information only, followed by a Tacotron-like model, conditioned on both speaker and accent IDs, which generates mel-spectrogram from the ASR-extracted features.
In our proposed approach, only one source spectrogram and its text transcription are needed during training and inference. Moreover, only the conversion model needs to be trained as opposed to the two-stage training approach of ~\cite{many-to-many-accent-conversion}. To the best of our knowledge, this is the first approach to satisfy both properties.

In this paper, we propose a novel AC approach based on normalizing flows~\cite{Kobyzev_2020,ho2019flow,kingma2018glow}.
Our model shares similarities with
other flow-based approaches applied to speech~\cite{flow_tts_9054484,kim2020glowtts,sc_glow_tts, flowtron, blow}, 
however, to the best of our knowledge, this is the first application of normalizing flows to the AC task. Moreover, we extend the flow-based conversion approaches to allow for conversion between source and target sequences of different lengths by using an attention mechanism, similar to~\cite{Seq2SeqVC}, and warping of speech phonemes' durations.

The main contributions of the paper are the following:
\begin{itemize}[noitemsep,nolistsep]
\item We propose a flow-based accent conversion model that significantly outperforms the competitive
  CopyCat baseline~\cite{Karlapati_2020}. The proposed model only requires a source utterance and its
  transcription and does not require the use of parallel data or a reference utterance at the inference stage.
\item We introduce a novel scheme for the AC task, referred to as 'remap, warp and attend'. The remap stage consists of remapping
  the conditioning phonetic sequence to match that of the target accent, the warp stage adjusts the durations to better suit the remapped phonemes while the attend stage learns to implicitly align the source and target speech signals whose phoneme sequences are of different lengths.

\end{itemize}

The remainder of the paper is organised as follows: Section 2 describes the proposed model, Section 3 presents the evaluation strategy used and discusses the results, while Section 4 presents the conclusions.

\section{Proposed approach}
\begin{figure*}[!ht]%
\centering
\subfloat[model architecture]{{\includegraphics[width=0.75\linewidth]{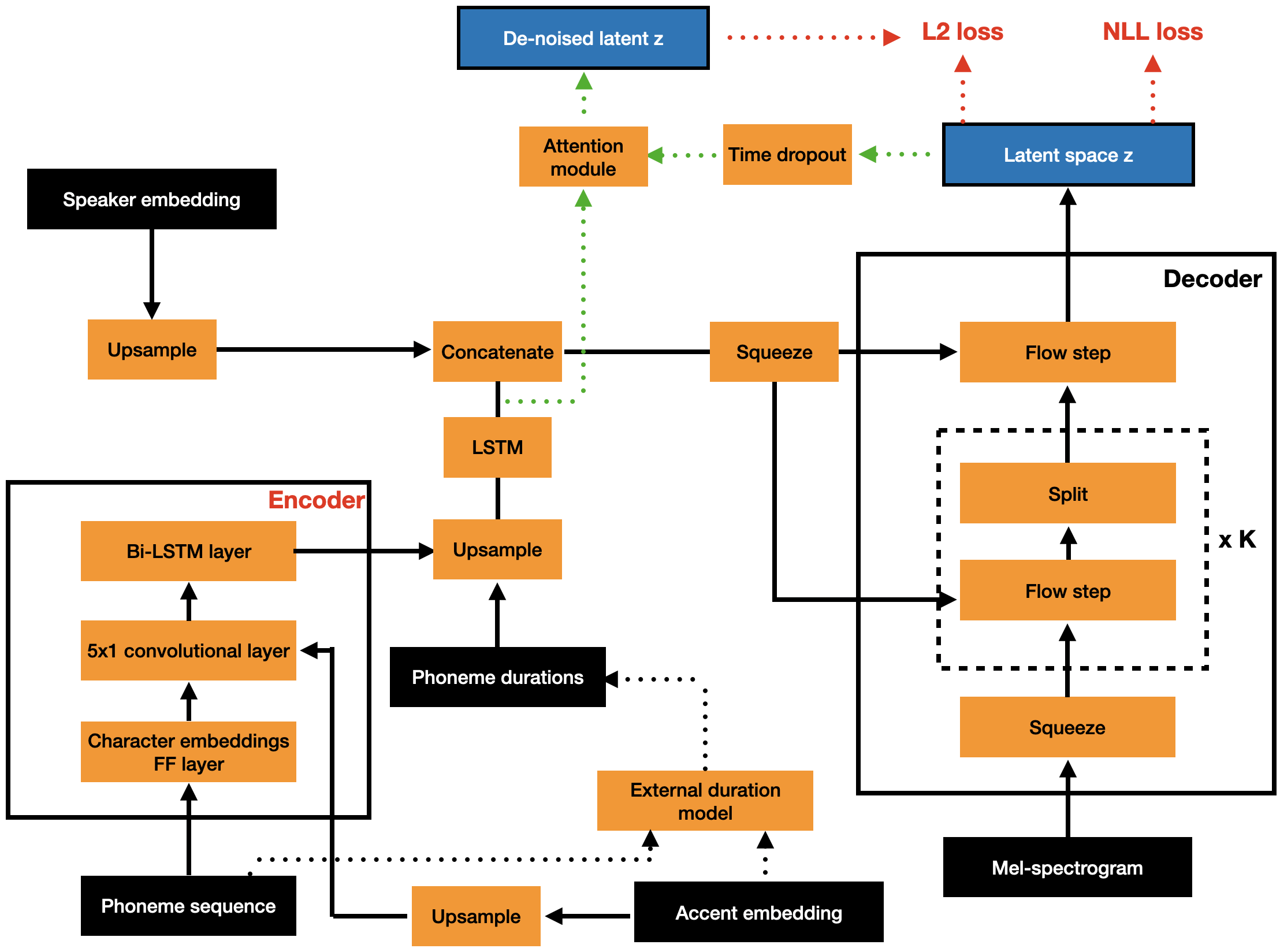}}}%
~~
\newline
\subfloat[Flow step]{{\includegraphics[width=0.3\linewidth]{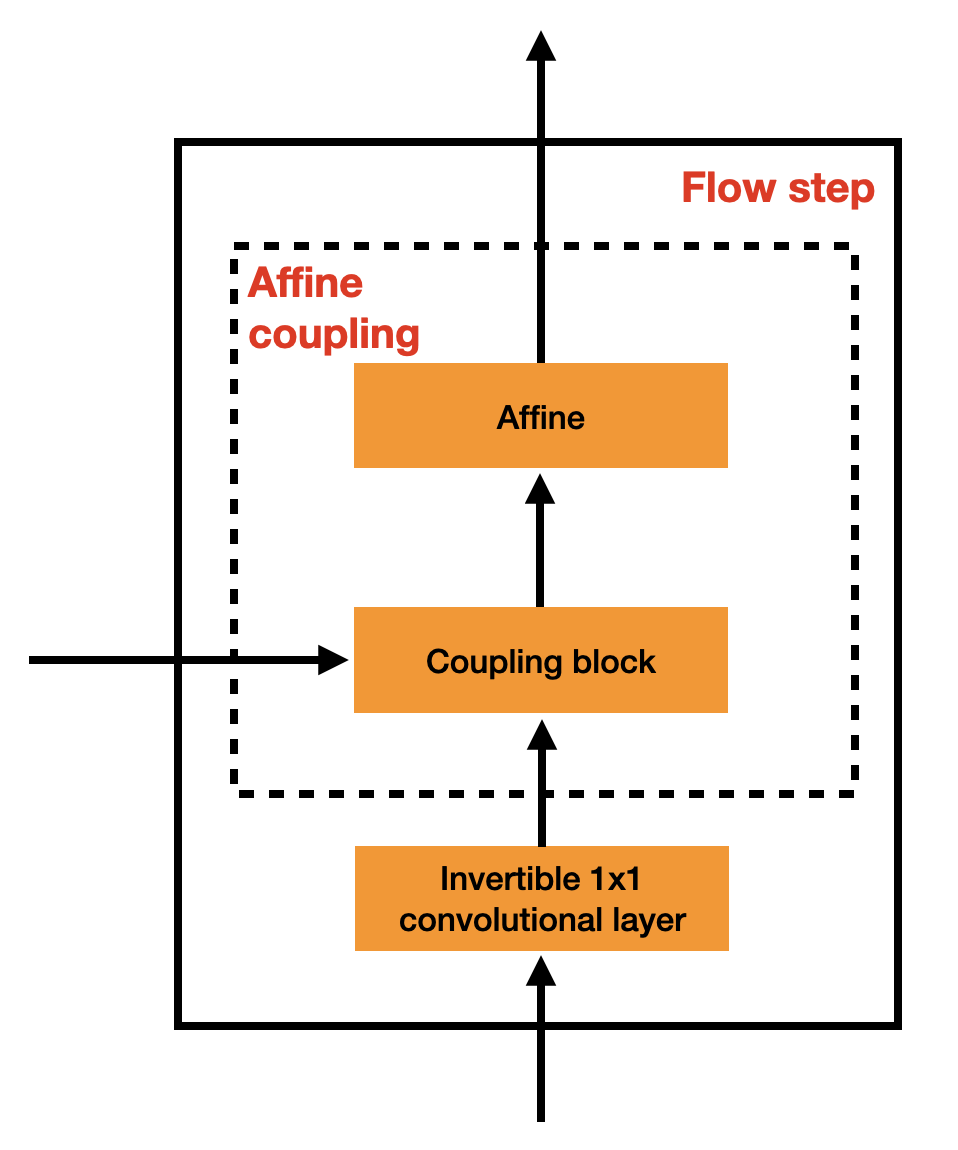}}}%
~~
\subfloat[Flow coupling block]{{\includegraphics[width = 0.3\linewidth]{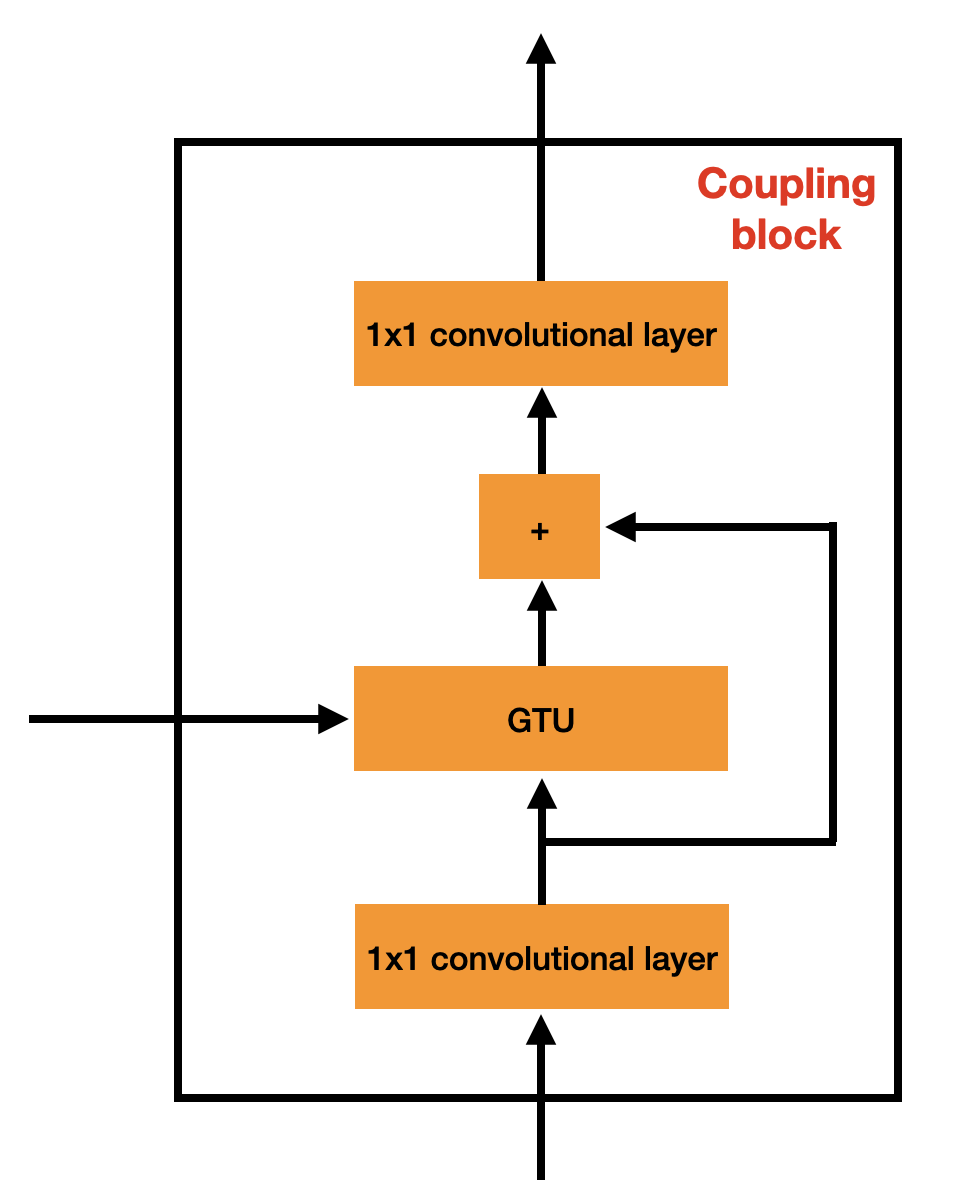}}}%
\caption{Overview of our accent conversion approach using normalizing flows (a) Flow step (b) and Coupling block (c). Arrows show the data flow during training. Black dotted arrows correspond to elements used with a duration model only, see Section.~\ref{sec:warp}. Green dotted arrows correspond to elements used with the attention block only, see Section.~\ref{sec:attend}. Red dotted arrows correspond to loss computation.}%
\label{fig:overview}%
\end{figure*}

\begin{figure*}
    \centering
    \includegraphics[width=0.75\linewidth]{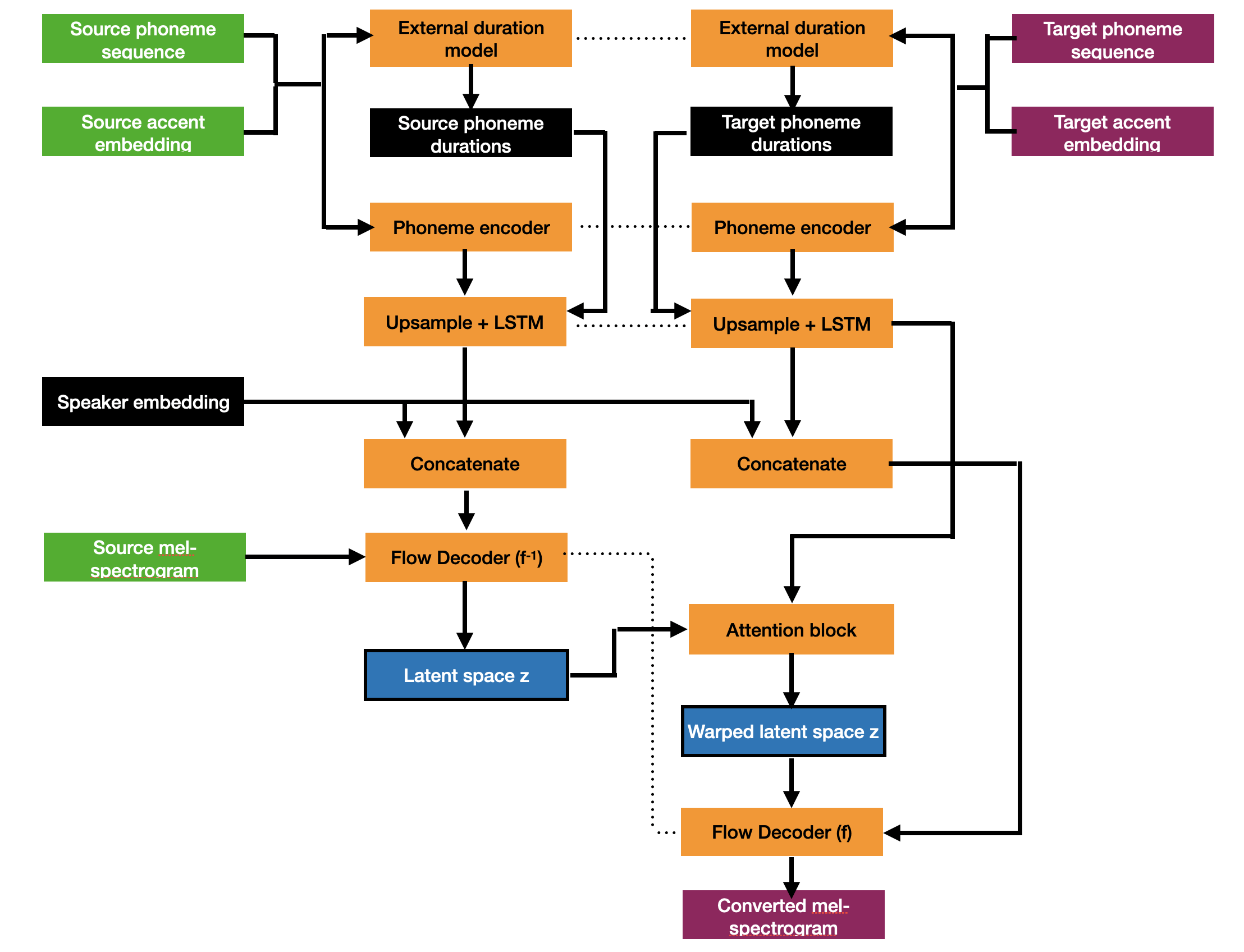}
    \caption{Accent conversion procedure. Dotted lines mean the modules linked are exactly the same and have the same weights.}
    \label{fig:data_flow}
\end{figure*}

Our approach is inspired by~\cite{Spring_VC} which used a Flow-TTS-like architecture~\cite{flow_tts_9054484} to achieve state-of-the-art results in the VC task. We extend this model in order to allow it to perform accent conversion. We achieve this goal by introducing a novel remap, warp and attend technique that significantly improves the quality of accent conversion.
Figure \ref{fig:overview} summarizes the architecture of the proposed model. Figures \ref{fig:data_flow} demonstrates the accent conversion procedure.

Our model is based on Normalizing Flows, and is able to encode a mel-spectrogram $x$ into a latent sequence $z$:
\begin{equation}
\label{sec2eq2}
  z = f^{-1}(x; ph, spk, acc),
\end{equation}
\noindent and then decode the latent sequence $z$ back into the original mel-spectrogram using the inverse transform $f$:
\begin{equation}
\label{sec2eq1}
  x = f(z; ph, spk, acc),
\end{equation}
where $ph$ is the phoneme sequence, $spk$ is the speaker embedding, and $acc$ is the accent embedding.
The Flow-based generative model explicitly learns the data distribution $p(x)$, and therefore it is optimized by minimizing the negative log-likelihood.
Similar to the VC model in~\cite{Spring_VC}, we use explicit durations, obtained using either a forced alignment~\cite{force_alignment} or predicted by a duration model, to upsample the phonetic representations, and we introduce utterance-level speaker and accent embeddings which are pre-trained as in~\cite{spk_acc_embeddings}. However, we chose not to condition the model on f0 and voiced/unvoiced flags in order to assess separately the effects of our proposed approach and prosody modelling on the AC's performance. Once we establish the effectiveness of our model for AC, we can extend it to incorporate f0 and voiced/unvoiced flags as it was the case for ~\cite{Spring_VC}.

%%%%%%%
%\begin{figure*}
%\centering
%\begin{subfigure}
%    \centering
%    \includegraphics[width=0.7\linewidth]{figures/model_architecture.png}
%    \caption{Image}\label{fig:image1}
%\end{subfigure}

%\newline

%\begin{subfigure}
%    \includegraphics[width=0.3\linewidth]{figures/Flow_layer.png}
%    \caption{Image}\label{fig:image1}
%\end{subfigure}

%\begin{subfigure}
%    \includegraphics[width=0.3\linewidth]{figures/coupling_block.png}
%    \caption{Image}\label{fig:image1}
%\end{subfigure}

%\caption{Overview of our accent conversion approach using normalizing flows (a) Flow step (b) and Coupling block (c). Black dotted arrows correspond to elements used with a duration model only, see Section.~\ref{sec:warp}. Green blocks and dotted arrows correspond to elements used with the attention block only, see Section.~\ref{sec:attend}.}
%\end{figure*}
%%%%%%

\subsection{Accent conversion via phoneme remapping 'remap'}
The Flow VC \cite{Spring_VC} model is extended to perform accent conversion. 
We propose to encode a mel-spectrogram $x_{source}$ into a latent sequence $z$ using a phoneme sequence $ph_{source}$ and an accent embedding $acc_{source}$ corresponding to the source accent:
\begin{equation}
\label{sec2seq3}
  z = f^{-1}(x_{source}; ph_{source}, spk_{source}, acc_{source}),
\end{equation}
\noindent and decode the latent sequence $z$ into a mel-spectrogram $x_{target}$ using a new phoneme sequence $ph_{target}$ and a  new accent embedding $acc_{target}$ corresponding to the target accent:
\begin{equation}
\label{sec2seq4}
  x_{converted} = f(z; ph_{target}, spk_{source}, acc_{target}),
\end{equation}
where $x_{converted}$ is the mel-spectrogram output of the conversion procedure and $spk_{source}$ is the speaker embedding corresponding to the source speaker. 

Accent embeddings help the model disambiguate how to pronounce phonemes that share the same symbol in different accents. 
The phoneme sequences $ph_{source}$ and $ph_{target}$ represent the same text but were extracted with two different grapheme-to-phoneme models, each dedicated to the source and target accents respectively. The `remap' stage is altering the phoneme sequence between the encoding and decoding stages to match the target accent (Equations~\ref{sec2seq3} and \ref{sec2seq4}).
%In order to confirm that phoneme remapping was required, we conducted an experiment where only the accent embedding changed between the encoding and decoding stages.
%Informal listening showed that no accent conversion was observed when the phoneme sequence was kept the same as the source one.
%Thus, we propose to alter both the phonemes and accent embedding to match the target accent during the decoding stage (Equations \ref{sec2eq3} and \ref{sec2eq4}).

\subsection{Duration warping 'warp'}
\label{sec:warp}
Remapping the phoneme sequences between the encoding and decoding stages improves accent control.
However, since we remap the source phoneme sequence into a target sequence without further manipulations, 
the durations of the target phonemes are still the same as those of the source sequence's.
We argue that it is possible to generate even more natural-sounding speech in the target accent by forcing 
both the phonemes and their durations to better match the target accent.

To achieve this, we propose duration warping: the goal of duration warping is to make
the encoded sequence of latent representations (whose length is $T_{source}$) have
the same length as speech in the target accent (the target speech length is denoted by $T_{target}$).\\
We first train a duration model. The duration model is
similar to the phonetic encoder as it uses the same inputs (phonemes and accent embedding) 
and has the same architecture. The only difference being that the duration model is trained separately using L2 loss.
Using the predicted durations, we create a warping matrix $W$ that will
serve to create a warped version of the latent sequence $z_{warped}$:
\begin{equation}
  z_{warped} = W \times z_{source},
\end{equation}
where $z_{source}$ is the sequence,  of length $T_{source}$, of latent representations encoded from the source mel-spectrogram,
$z_{warped}$ is the warped sequence of representations that will be used at the decoding stage (of length $T_{target}$),
and $W$ is the warping matrix of size $T_{target} \times T_{source}$. Multiplying $z_{source}$ with $W$
has the same effect as applying interpolation to $z_{source}$ (similar to using \textit{interpolate}
in common deep learning frameworks) but with a scaling factor that varies across phonemes instead of being fixed
for the whole utterance.

The accent conversion with duration warping can be summarized as running the same encoding stage as in Equation~\ref{sec2seq3},
but then replacing $z$ by $z_{warped}$ in Equation~\ref{sec2seq4} during the decoding stage.

\subsection{Improving duration warping with attention 'attend'}
\label{sec:attend}
The combination of phoneme sequence remapping and duration warping leads to a better
control of accent.
However, such approach has one main limitation: constructing the warping matrix requires access to an explicit alignment between the source and target phoneme sequences which is then used to align the target phonemes $ph_{target}$
with the encoded latent representations $z_{source}$.

In order to circumvent the above limitation, we introduce an attention block which takes
a phoneme sequence as queries and the sequence of latent representations as both keys and
values. The idea behind the attention block is to enable the model to learn an implicit
alignment between the target sequence of warped latent representations and the source sequence of representations, encoded from the source speech signal, without the need for an explicit definition of a warping matrix. The attention can also be seen as a more powerful generalization of interpolating using a warping matrix $W$.

%During the training phase, the attention
%block is trained using L2 loss to reconstruct sequence of source latent representations from both
%the sequence of phonemes and a noisy version of the source latent representations themselves.
%We use a noisy version of the source latent representations as input to avoid the attention
%collapsing into a point where it learns an identity transformation between its input and output
%latent representations. We generate the noisy version of the source latent representations by applying
%time-dropout where a percentage of frames are zeroed-out at training time.

During the training phase, the attention block trained using L2 loss to denoise a noisy version of the sequence of the source latent representations. We use a noisy version of the source latent representations as input to avoid the attention collapsing into a point where it learns an identity transformation between its input and output latent representations. We generate the noisy version of the source latent representations by applying time-dropout where a percentage of frames are zeroed-out at training time.

\section{Experiments and Results}
We evaluate the performance of the proposed approach in terms of intelligibility, accent similarity, naturalness and speaker similarity. 

We chose CopyCat as a baseline model. CopyCat showed state-of-the-art performance in the Voice Conversion task. We adapt CopyCat to the AC task by conditioning the model on speaker and accent embeddings where both embeddings are concatenated to the upsampled phoneme sequence before being passed to the phoneme encoder, which is in line with~\cite{copycat_changes}. We also applied phoneme remapping and duration warping to the CopyCat baseline.

\subsection{Dataset}
To evaluate accent conversion, we train our flow model and the CopyCat baseline
on a multi-speaker, multi-accent proprietary dataset. The dataset comprises 3173 speakers distributed
across 6 English accents: American, Australian, British, Canadian, Indian and Welsh. Each accent 
has a different number of utterances, ranging from around 18k to 300k. The number of utterances also 
differs across speakers, ranging from 100 to 20k. The recording conditions vary across the dataset, with some speakers recorded in studio quality conditions
whilst other speakers were recorded with lower quality microphones in more ambient surroundings.

\subsection{Perceptual evaluations setup}
All perceptual evaluations were crowdsourced using the same settings unless explicitly stated otherwise. The evaluations were conducted on the conversion from a male American English speaker to the 5 remaining accents: Australian, British, Canadian, Indian and Welsh. A test set of 25 utterances was selected, resulting in 125 effective utterances for evaluation (25 utterances * target accents). The utterance lengths ranged from 4 words to 25 words. All audio samples were vocoded using a universal vocoder \cite{universal_vocoder} except for recordings.
The evaluations were conducted with 300 listeners, each rating 25 utterances. 
To test for statistically significant differences between systems we perform paired t-tests. Holm-Bonferroni correction is applied due to the number of systems being compared. 
In the remainder of this paper, we use the term \textit{statistically significant} to refer
to $p \leq  0.05$.

It is also worth noting that the proposed AC approach of remapping the phoneme sequence to the target accent, 
relies on the assumption that the target phoneme sequence can still be aligned with the 
source mel-spectrogram to be converted. This alignment is implicitly learned by our final model: \textit{remap-warp-attend}.
However, the \textit{remap-warp} model and the \textit{CopyCat} baseline need an explicit alignment
which requires aligning the source and target phoneme sequences. We simplify the evaluation 
by restricting the test set to utterances for which $ph_{source}$ and $ph_{target}$ have the same number of phonemes.
This evaluation restriction allows us to perform an ablation study of the components without the need to externally align source and target phoneme sequences. We note that \textit{remap-warp-attend}
can handle any general case and doesn't need such restriction to be evaluated.

\subsection{Intelligibility}
We measure the intelligibility of the models using Word Error Rate (WER). WER was computed by comparing the original text of an utterance with the transcription, obtained using AWS Transcribe ASR system, of the converted speech. 

WER results summarized in Table~\ref{tab:obj_eval} show that speech generated by the proposed approaches
(remap-warp and remap-warp-attend) are signiﬁcantly more intelligible than other models, with up to 56\% reduction in WER
compared to the CopyCat baseline. As such it can be seen that
remap-warp provides a more stable conversion compared to the
baseline and that the attention block in remap-warp-attend allows the ﬂow model to handle the general
accent conversion case without loss to the intelligibility of the
converted speech. Remap had a sub-par performance compared
to the other ﬂow models in terms of intelligibility. We hypothesize that this performance is explained by the poor match between the source durations and the target phoneme sequence to be upsampled. The remap model naively assumes that all
changes between source and target phoneme sequences are the
result of phoneme substitutions if both sequences are of the
same length. However, it is likely that insertions and deletions
of phonemes did occur which is not handled by said model.

\begin{table}[h!]
    \begin{center}
    \caption{\footnotesize{WER, relative WER and WERR (WER Reduction) per accent conversion system baselined against CopyCat.}}
    \label{tab:obj_eval}
    \scalebox{1.0}{
    \resizebox{\columnwidth}{!}{%
    \begin{tabular}{l l l}
    \hline
    %\multicolumn{2}{c}{\textbf{Relative WER (and WERR)}} 
    System & WER & Relative WER (WERR)\\
    \hline
    \textit{CopyCat} & 26.86\% $\pm$ 2.74 & 1 \\
    \textit{Remap} & 20.40\% $\pm$ 2.49 & 0.76 (24\%)\\
    \textit{Remap-warp} & 13.43\% $\pm$ 2.11 & 0.50 (50\%)\\
    \textbf{\textit{Remap-warp-attend}} & \textbf{11.84\%} $\pm$ \textbf{2.00}& \textbf{0.44 (56\%)}\\
    \hline
    \end{tabular}}
    }
    \end{center}
\end{table}

\subsection{Accent similarity: duration warping effect}
We assess the impact of duration warping on the accent conversion using a MUSHRA test. For each MUSHRA screen, listeners listen to a reference recording from a different speaker with the same gender speaking in the target accent. The testers were then asked to ``\textit{rate how similar the accents in each system sounds compared to the reference speaker}''.

Results, summarized in Table~\ref{tab:subj_eval_acc_sim_warp}, conﬁrm signiﬁcant changes from the source accent
towards the target accent for all proposed models. All the differences between the systems in this evaluation were signiﬁcant
except for the difference between remap-warp and remap-warp-attend. The results from this evaluation suggest that both the
remap and warp steps played an important role towards achieving better accent conversion by making the phonemes and their duration better match the target accent. The attend step allows the model
to handle the general conversion case, where source and target
phoneme sequences are not explicitly aligned, without degrading the accent similarity score.

\begin{table}
    \caption{\footnotesize{Average scores for the accent similarity evaluation which assesses the effect of duration warping.}}

    \label{tab:subj_eval_acc_sim_warp}
    \begin{center}
    \scalebox{1.0}{
    \begin{tabular}{l c}
    \hline
    \multicolumn{2}{c}{\textbf{Accent similarity - duration warping effect}} \\ \hline
    \textit{Source speaker recordings - lower anchor} & 57.53\\
    \textit{Target accent recordings - upper anchor} & 68.57\\
    \textit{Remap} & 60.41\\
    \textbf{\textit{Remap-warp}} & \textbf{62.13}\\
    \textit{Remap-warp-attend} & 61.51\\
    \hline
    \end{tabular}}
    \end{center}
 \end{table}
 
 \subsection{Accent similarity: baselining}
 We compare our flow approaches against CopyCat using a MUSHRA test where we use the same setup as in section 3.4. Results of this evaluation are summarized in Table~\ref{tab:subj_eval_acc_sim_baseline}.

All differences
between systems in the above evaluation are signiﬁcant except
for the difference between remap-warp and remap-warp-attend. These results show that the proposed ﬂow-based approach signiﬁcantly improves AC accuracy, with both remap-warp and remap-warp-attend signiﬁcantly outperforming CopyCat. A potential explanation to the previous observation, is that the lossless nature of flow-models, where they are able to exactly reconstruct their input mel-spectrogram if the conditionings stay the same, is more beneficial to the AC task than the lossy representation of speech that CopyCat uses, where speech is decoded after using a time bottleneck. The results also conﬁrm the ﬁndings from the previous accent similarity test
indicating that the attention block adds ﬂexibility to ﬂow models
without degrading the accent similarity scores.

\begin{table}
    \caption{\footnotesize{Average scores for the accent similarity evaluation which baselines our flow model against CopyCat.}}

    \label{tab:subj_eval_acc_sim_baseline}
    \begin{center}
    \scalebox{1.0}{
    \begin{tabular}{l c}
    \hline
    \multicolumn{2}{c}{\textbf{Accent similarity - baseline}} \\
    \hline
    \textit{Source speaker recordings - lower anchor} & 62.14\\
    \textit{Target accent recordings - upper anchor} & 71.44\\
    \textit{CopyCat} & 63.05\\
    \textit{Remap-warp} & 65.26\\
    \textbf{\textit{Remap-warp-attend}} & \textbf{65.31}\\
    \hline
    \end{tabular}}
    \end{center}
 \end{table}

\subsection{Naturalness}
After establishing the proposed model's capabilities in terms of accent control, we evaluate how natural-sounding is its synthesized speech.
To this end, we measure the naturalness of the proposed approach using a MUSHRA test. The testers were asked to ``\textit{rate the audio samples in terms of their naturalness}''. Results are summarized in Table~\ref{tab:subj_eval_naturalness}.

All the differences were signiﬁcant except the differences between remap-warp and remap-warp-attend.
The results suggest that both remap-warp and remap-warp-attend are the most
natural-sounding AC systems surpassing the CopyCat baseline. 
The results also show that the
attention block doesn’t degrade the naturalness while still allowing more ﬂexibility to the AC ﬂow model. This is in-line with similar observations made previously in the accent similarity evaluations.

\begin{table}
    \caption{\footnotesize{Average scores for naturalness evaluations comparing accent conversion approaches.}}

    \label{tab:subj_eval_naturalness}
    \begin{center}
    \scalebox{1.0}{
    \begin{tabular}{l c}
    \hline
    \multicolumn{2}{c}{\textbf{Naturalness - baseline}} \\
    \hline
    \textit{Source speaker recordings - upper anchor} & 75.30\\
    \textit{CopyCat} & 68.15\\
    \textit{Remap-warp} & 69.11\\
    \textbf{\textit{Remap-warp-attend}} & \textbf{69.62}\\
    \hline
    \end{tabular}}
    \end{center}
 \end{table}

\subsection{Speaker similarity}
The focus of the model is to perform accent conversion while leaving the perceived speaker identity unchanged. We measure potential alteration to the speaker identity using a MUSHRA test. Testers are presented with a recording from the source speaker and are then asked to ``\textit{rate how similar the speakers in each system sound compared to the reference speaker}". The reference audio was always a different text content from the text of the samples evaluated. 

Results from this evaluation are summarized in Table~\ref{tab:subj_eval_spk_sim}. All differences in the evaluation are signiﬁcant. All AC systems show some perceived changes to the speaker identity. The
signiﬁcant difference between remap-warp and remap-warp-attend suggests that the attention mechanism does alter the perceived speaker identity. Following informal listening, we hypothesise that this may be due to a tendency of the attention to flatten the prosody of converted speech. However,
more investigation is required to better understand how the
prosody changes may have impacted the speaker similarity
test's results without impacting the accent similarity. CopyCat seems to alter the speaker identity the least. This
observation may be explained by the fact that CopyCat makes
smaller changes to the source speech, as suggested by its lower
accent similarity scores, and is thus more likely to restore the
source speaker’s identity in the converted speech.

\begin{table}
    \caption{\footnotesize{Average scores for speaker similarity evaluation comparing accent conversion approaches.}}

    \label{tab:subj_eval_spk_sim}
    \begin{center}
    \scalebox{1.0}{
    \begin{tabular}{l c}
    \hline
    \multicolumn{2}{c}{\textbf{Speaker similarity - baseline}} \\
    \hline
    \textit{Target accent recordings - lower anchor} & 49.11\\
    \textit{Source speaker recordings - upper anchor} & 75.67\\
    \textbf{\textit{CopyCat}} & \textbf{63.53}\\
    \textit{Remap-warp} & 62.53\\
    \textit{Remap-warp-attend} & 61.06\\
    \hline
    \end{tabular}}
    \end{center}
 \end{table}

\subsection{Many-to-many accent conversion}
To confirm the proposed approach's AC performance is stable across all combinations of source and target accents, we perform further evaluations. The MUSHRA test has the same setup as in section 3.4 except for the selection of the test set: one male and one female speakers were selected per source accent, we then select 27 utterances per speaker and run accent conversion towards all remaining target accents. The only exception is the Canadian English accent for which we selected only one female speaker. This setup yields an effective test set of 1485 evaluated utterances (27 utterances * 11 speakers * 5 target accents). Each evaluation screen showed the following systems: reference recordings in the target accent, our remap-warp-attend samples, source speaker recordings (lower anchor), and recordings from a speaker speaking the target accent who is different from both the source and reference speakers (upper anchor).

We report in Table~\ref{tab:many_to_many_ratio} the ratio between the means of the MUSHRA scores of our remap-warp-attend model and the target accent recordings for each source-target accent pair. We chose to report the ratio between our system's and the upper anchor's MUSHRA scores instead of the absolute numbers since the upper anchors scored differently in each accent. We omit results on conversion to and from Indian English as we noticed later in the development process some labelling issues with the Indian English data. We confirmed after fixing the issue that the other accents were not affected.

Overall, we observe consistently high accent similarity scores ratios. It is worth noting that in multiple conversion combinations, the average accent similarity rating for the proposed approach achieved a ratio close to, or higher than, 1. This can be interpreted as the listeners not being able to distinguish the accent of converted speech from that of recordings of a native speaker of that accent. The observed ratios are lower for conversion to Canadian English compared to other locales. This is likely because the target accent speaker used for the reference was the same speaker as the target accent speaker used as upper anchor for this accent. This resulted in Canadian English upper anchor being scored much higher in comparison to the other accents'.

From the many-to-many conversion results, we conclude that the proposed approach is capable of converting accent across all seen combinations of source/target accents.

\begin{table}[H]
    \caption{\footnotesize{Matrix summarizing ratios of accent similarity scores between remap-warp-attend and target recordings for the conversion from source accents (rows) to target accents (columns).}}
    \label{tab:many_to_many_ratio}
    \begin{center}
    \scalebox{1.0}{
    \begin{tabular}{l l l l l l l}
    \hline
     & en\_au & en\_ca & en\_gb  & en\_us & en\_wls\\ 
     en\_au & - & 0.66 & 0.94 & 0.99 & 0.98\\
     en\_ca & 0.86 & - & 0.96 & 0.98 & 0.87\\
     en\_gb & 0.97 & 0.69 & - & 0.97 & 0.99 \\
     en\_us & 0.95 & 0.66 & 1.00 & - & 0.91\\
     en\_wls & 1.04 &  0.64 & 0.98 & 0.98 & -\\
    
    \hline
    \end{tabular}}
    \end{center}
 \end{table}

\section{Conclusion}
We propose an accent conversion method based on an innovative extension
to normalizing flows called 'remap, warp and attend'. The remap step
converts the sequence of phoneme conditioning between source and target
speech to match the target accent. The warp step adjusts the
phoneme durations, enabling conversion between source and target speech
signals of different durations. Finally, the attend step allows the model to handle the general AC case, by removing constraints on its input sequences, while maintaining the same naturalness and accent similarity scores. Through objective and
perceptual evaluations, we show that the approach significantly improves
conversion compared to the CopyCat baseline in terms of
accent similarity, naturalness and intelligibility. While the proposed approach is
tailored to accent conversion, the “remap, warp and attend” technique
paves the way for conversion of multiple speech attributes, where any
speech property can be independently adjusted.

\bibliographystyle{IEEEtran}

\bibliography{mybib}

% \begin{thebibliography}{9}
% \bibitem[1]{Davis80-COP}
%   S.\ B.\ Davis and P.\ Mermelstein,
%   ``Comparison of parametric representation for monosyllabic word recognition in continuously spoken sentences,''
%   \textit{IEEE Transactions on Acoustics, Speech and Signal Processing}, vol.~28, no.~4, pp.~357--366, 1980.
% \bibitem[2]{Rabiner89-ATO}
%   L.\ R.\ Rabiner,
%   ``A tutorial on hidden Markov models and selected applications in speech recognition,''
%   \textit{Proceedings of the IEEE}, vol.~77, no.~2, pp.~257-286, 1989.
% \bibitem[3]{Hastie09-TEO}
%   T.\ Hastie, R.\ Tibshirani, and J.\ Friedman,
%   \textit{The Elements of Statistical Learning -- Data Mining, Inference, and Prediction}.
%   New York: Springer, 2009.
% \bibitem[4]{YourName17-XXX}
%   F.\ Lastname1, F.\ Lastname2, and F.\ Lastname3,
%   ``Title of your INTERSPEECH 2022 publication,''
%   in \textit{Interspeech 2022 -- 23\textsuperscript{rd} Annual Conference of the International Speech Communication Association, September 18-22, Incheon, Korea, Proceedings, Proceedings}, 2022, pp.~100--104.
% \end{thebibliography}

\end{document}